\documentclass[fp]{jpsj3}
\usepackage{graphicx}
\begin{document}
\title{Temperature Dependence of Highly Excited Exciton Polaritons in Semiconductor Microcavities}

\author{Tomoyuki Horikiri$^{1,2,3,4}$, Yasuhiro Matsuo$^{1,3}$, Yutaka Shikano$^{5,6}$, Andreas L\"{o}ffler$^7$, Sven H\"{o}fling$^{1,7}$, Alfred Forchel$^7$, and Yoshihisa Yamamoto$^{1,2,3,4}$}

\inst{$^1$National Institute of Informatics, 2-1-2 Hitotsubashi, Chiyoda-ku, Tokyo 101-8430, Japan\\
$^2$E. L. Ginzton Laboratory, Stanford University, Stanford, CA 94305, USA\\
$^3$The University of Tokyo, 7-3-1 Hongo, Bunkyo-ku, Tokyo 113-8656, Japan\\
$^4$Center for Emergent Matter Science, RIKEN, Wakoshi, Saitama 351-0198, Japan \\
$^5$Research Center of Integrative Molecular Systems (CIMoS), Institute for Molecular Science, 38 Nishigo-Naka, Myodaiji, Okazaki 444-8585, Japan \\
$^6$Schmid College of Science and Technology, Chapman University, Orange, CA 92866, USA \\
$^6$Technische Physik, Universit\"{a}t W\"{u}rzburg, Am Hubland, D-97074 W\"{u}rzburg, Germany}

\abst{Observations of polariton condensation in semiconductor
microcavities suggest that polaritons can be exploited as a novel
type of laser with low input-power requirements. The low-excitation
regime is approximately equivalent to thermal equilibrium, and a
higher excitation results in more dominant nonequilibrium features.
Although standard photon lasing has been experimentally observed in the 
high excitation regime, e-h pair binding can still remain even in 
the high-excitation regime theoretically. Therefore, the photoluminescence with 
a different photon lasing mechanism is predicted to be different from that with a
standard photon lasing. In this paper, we report the temperature dependence 
of the change in photoluminescence with the excitation density. The second 
threshold behavior transited to the standard photon lasing is not measured at 
a low-temperature, high-excitation power regime. Our results suggest that 
there may still be an electron--hole pair at this regime to give a different photon 
lasing mechanism.}

\kword{polariton, photon lasing, semiconductor, microcavity, Bose-Einstein condensation} \maketitle

\section{Introduction}
Bosonic quasiparticles resulting from the strong coupling between
photons and excitons are known as polaritons. Exciton polaritons are
manifested by the spectroscopic observation of the photoluminescence (PL) 
from semiconductor microcavities. When the rate of inflow to
the ground state of polaritons exceeds the rate of loss from the
cavity, final-state stimulation can result in a macroscopic
population of the ground state, and ground-state polariton
condensation can occur \cite{snoke,huireview, sch1}. This defines
the condensation threshold, which can be experimentally confirmed by
a nonlinear increase in the PL from a semiconductor microcavity 
as a function of the input pump power, as
polaritons inside the cavity cannot be observed directly. Owing to
the short lifetime of polaritons, this condensation is treated as a
nonequilibrium dynamical condensation \cite{hui}; however, under
certain conditions, such as polariton Bose-Einstein condensation
(BEC), it may be treated as a thermal equilibrium condensation
\cite{kasprzak1}.

There is a large body of experimental evidence that suggests
that polariton condensation is a weakly interacting BEC induced by the
polariton--polariton interaction. For example, observations of
superfluidity~\cite{amo}, vortex formation~\cite{lagoudakis},
vortex--antivortex pair creation~\cite{george}, and the spectrum of
the Bogoliubov excitation~\cite{utsunomiya} all support this
hypothesis. Furthermore, polariton condensation has the potential to
be exploited for developing a novel type of laser with low input-power
requirements~\cite{imamoglu}. Unlike standard photon lasing in a
semiconductor, population inversion is not needed, and an increase
in the pump power leads to an increase in the strength of
polariton--polariton interaction effects owing to the population
increase. At very high densities, exciton overlap screens the
Coulomb interaction, which breaks electron--hole (e--h) pairs and
results in an e--h plasma system. Population inversion then occurs,
and photon lasing is observed after the second nonlinear increase,
which is often called the second threshold in contrast to the first
or condensation
threshold~\cite{dang,hui2,bajoni,nelsen,bajoni2,elena,tempel,tempel2}.
However, recent theoretical
studies~\cite{kamide1,tim,kamide2,yamaguchi,yamaguchi2} have
indicated that some e--h pairs can remain in the high-density
regime.

The aim of this study is to clarify whether the high excitation
regime can be detected in the PL behavior, as direct observation of
e--h pairs and the e--h plasma is not possible. To compare the PL behavior 
in the high-excitation regime with standard photon lasing, we investigated the
temperature dependence of the PL because the high-temperature case
shows standard photon lasing. Here, we demonstrate a completely
different PL behavior from that of standard photon lasing at a low
temperature ($8$ K) in the high-excitation regime.

In \S~\ref{review}, we present the basic concepts of semiconductor
microcavity polaritons and review recent developments in the
high-density regime in \S~\ref{high_dens}. In
\S~\ref{our_result}, the results of the excitation dependence
measurements, in which the PL intensity, energy, PL
linewidth, and second-order correlation function are studied as functions 
of pump power, are presented. 
These results are discussed with regard to temperature to compare the 
condensation and lasing mechanisms in \S~\ref{discussion}. 
Furthermore, we evaluate the excitation
density of the e--h pairs in our experimental setup to identify the
high-excitation regime for which the PL behavior is different from
standard photon lasing. Section~\ref{conclusion} is devoted to a
summary.

\section{Polariton Condensation and Photon Lasing \label{review}}
\begin{figure*}[t]
   \begin{center}
    \scalebox{0.8}{
    \includegraphics{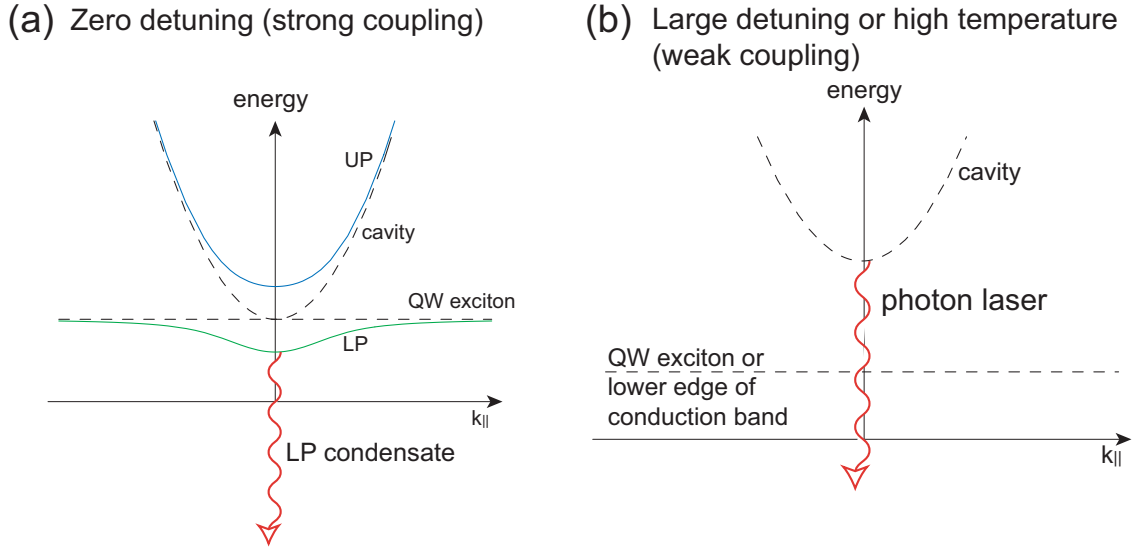}}
    \caption{(Color Online) Energy--momentum relationships for the strong and weak
coupling regimes. (a) Polariton condensation. (b) Photon lasing, for
which population inversion is necessary.}\label{1}
   \end{center}
\end{figure*}

Figure \ref{1} (a) shows that, in a semiconductor microcavity with a
quantum well (QW), a polariton is created by the strong coupling
between the QW exciton and a cavity photon, with the exciton itself
formed by e--h binding. As a result of fast energy transfer between
two cavity modes, normal-mode splitting occurs, and two energy
branches referred to as the upper polariton (UP) and lower polariton
(LP) are formed.
When the conditions in the cavity meet those of the condensation
threshold, a polariton condensate is formed in the LP ground state,
that is, when the rate of injection into the LP ground state through
relaxation by e--h pair, e--h and phonon, polariton--polariton, and
polariton--phonon scattering is higher than the rate of loss from the
ground state owing to inverse processes and cavity leakage. The
macroscopic population in the LP ground state then grows, and a
nonlinear increase in the PL is observed. This condensate is often
called the polariton laser. The difference between this and photon
lasing is that the strong PL emission is not caused by a conventional 
lasing mechanism, i.e., light amplification by stimulated emission
of radiation.

Conventional semiconductor photon lasing in a photon laser, as shown
in Fig.~\ref{1} (b), occurs when there is no longer any strong coupling
between the cavity photon and the QW exciton. At room temperature,
excitons dissociate owing to the high thermal energy, and since the
exciton level no longer exists in a standard semiconductor, such as GaAs, 
the cavity energy in the conduction band is that of an e--h plasma.
Photon lasing relies on population inversion with respect to the
ground state, but there are also other differences between polariton
condensation and photon lasing.

The normalized second-order correlation function $g^{(2)}(0)$ of
photon lasing converges to unity just above the lasing threshold
\cite{ulrich}, while that of the polariton condensate shows bunching
($g^{(2)}(0)>1$) and super-Poisson statistics above the condensation
threshold \cite{hui,lausanne}. This can be understood as being a
result of the interactions between condensate polaritons or between
a polariton and phonon, which deplete the condensate population as
polaritons are lost from the cavity as a result of these
interactions \cite{sch1,sch2,haug,wouters2,haug2}. Owing to this
mechanism, higher-order coherence of the condensate is not realized,
and in fact, the third-order correlation function $g^{(3)}(0)$ has
been shown to be larger ($g^{(3)}(0)>g^{(2)}(0)>1$) \cite{horikiri}.
The coherent state exhibits the all-order coherence $g^{(n)}(0)=1$,
while the thermal state exhibits $g^{(3)}(0)=6
>g^{(2)}(0)=2 >1$. Consequently, the statistics of the PL from the
polariton condensate indicate that the condensate is not in the
coherent state but close to the thermal distribution state.

\begin{table*}[ht]
\begin{center}
\caption{Features of polariton condensation and photon lasing. The
cavity photon energy is $\omega_{\rm cav}$, $\omega_{\rm LP}$ is the
LP ground state energy, and $\mu$ is the condensate blue shift due to
the interactions.} \footnotesize\rm
\begin{tabular}{cccc} \hline
&$g^{(2)}(0)$&Energy level &Linewidth \\ \hline Polariton condensate
&$>1$ & $\omega_{\rm LP}+\mu$ &Broadening owing to p--p interaction\\
Photon laser&$\rightarrow 1$&$\omega_{\rm cav}$&Narrowing owing to
cavity linewidth \\ \hline
\end{tabular}
\label{2}
\end{center}
\end{table*}
The differences between polariton condensation and photon lasing are
summarized in Table \ref{2}. Differences in the energy
and linewidths, which are observed in the PL measurements, also arise. In
polariton condensation, the PL energy differs from that of the LP
ground state ($\omega_{\rm LP}$) by a small blue shift due to
polariton interactions ($\mu$), while the energy of the photon laser
is the same as the cavity photon energy of zero in-plane momentum
($\omega_{\rm cav}$). The polariton linewidth broadens as the
density increases owing to the polariton--polariton (p--p)
interactions \cite{porras}, while the photon laser shows a narrow
linewidth owing to their absence.

\section{Behavior in a High Excitation Regime \label{high_dens}}
When a polariton system reaches the condensation regime above the
condensation threshold, a transition to photon lasing in a
higher-excitation regime is considered to occur
\cite{dang,hui2,bajoni,nelsen,bajoni2,tsotsis,elena,tempel,tempel2}.
The PL intensity has a second threshold \cite{dang,bajoni2,tsotsis},
where population inversion is believed to be the reason for the
transition to photon lasing. In ref.~\citen{elena}, a study
of the Bernard--Duraffourg condition, which defines photon lasing
conditions in semiconductors, revealed that the system was in a
regime close to the inversion threshold at the onset of photon
lasing. The observation of the second-order coherence function
$g^{(2)}(0)$ has interestingly shown that it monotonically converges
to unity \cite{tempel,tempel2} above the photon lasing threshold.
The results of these studies can be understood as follows. With an
increase in the population, e--h binding in the high excitation
regime weakens owing to the screening effect between e--h pairs, and
the excitons become an e--h plasma. Consequently, standard photon
lasing takes place after the gain for the cavity photon mode exceeds
its loss. However, it is unclear whether the breaking of e--h pairs,
determined by the e--h interaction and e--h density, necessarily
occurs in every microcavity QW sample. The polariton condensate is
expected to experience several phases governed by the excitation
density and detuning, for which different signatures of the photon
laser have been theoretically determined under suitable conditions
\cite{kamide1,tim,kamide2,yamaguchi,yamaguchi2}. In particular, the
work described in refs.~\citen{yamaguchi} and \citen{yamaguchi2} utilized the
framework of a nonequilibrium system caused by pumping and decay
\cite{szymanska}, which is very close to the actual experimental
conditions. In the high-density regime, e--h binding is no longer a
result of the Coulomb interaction; instead, photon induced
attraction takes place. As a result, a unique dip in the momentum
distribution (kinetic hole burning) appears~\cite{yamaguchi}.

In these studies, in which a transition to photon lasing in the
high-density regime was observed, polaritons no longer survived
after the transition; excitons dissociated and formed an e--h plasma
in the high-excitation regime. However, theoretical studies predict
that another regime that does not necessarily experience e--h pair
breaking should exist in the high-density regime. There are
currently no experimental conditions that are known to implement
e--h pair binding at high densities; however, by exploring different
Q-factors, detuning values, excitation properties (angle, spot size,
and energy), and a lower heating effect from the pumping laser, it
may be possible to observe differences in the PL behavior from those
of photon lasing. In several studies, the LP resonance has been
utilized, but there is some disagreement over whether the coherence
from the pumping laser can be transferred to the polariton
condensate since the energy is close to the pumping energy and
populated within a few scattering events. Therefore, the pumping
energy (UP or cavity resonance at a large momentum) was chosen for
this work in such a way that a high input rate and a high excitation
density were obtained. To do so, the angle needs to be tuned to
the reflection dip of the sample. To guarantee a sufficiently
high energy, the reflection dip of the stop band edge of a
distributed Bragg reflector (DBR) structure was used because we were
not able to operate the laser stably at high energies. In terms of
the spot size, if normal pumping was performed, then a clearer,
smaller pumping for the condensate spatial profile might be
obtained, but this was not possible owing to the limited pumping
energy of the sample. Previous studies have considered samples with
different Q-factors, but the unique behavior we report here was only
obtained in the DBR sample we used. Thus, in what follows, we
will examine the trends in the PL behavior on the basis of the temperature and
excitation density of the sample. We find that a transition to
photon lasing does not completely explain the results at low
temperatures, which imply that the effects of e--h pairs are still
present in the high-excitation regime.

\section{Temperature and Excitation Power Dependences of PL \label{our_result}}
The low-temperature condition in which a polariton condensate can be
formed may show a different behavior from that of a photon laser, and
to compare the results of the temperature and excitation density
dependences of the PL\cite{tempel2} with the photon lasing case, we
also present high-temperature measurements for which e--h pairs
dissociate owing to the thermal energy and the cavity photon mode is
in the conduction band. Moreover, additional experiments at
intermediate temperatures also show an interesting behavior, including
the appearance of the second threshold of the transition from a
polariton condensate to a photon laser, which highlights the absence
of the second threshold of the PL intensity at low temperatures.

\subsection{Experimental setup}
\begin{figure}
   \begin{center}
    \scalebox{0.6}{
    \includegraphics{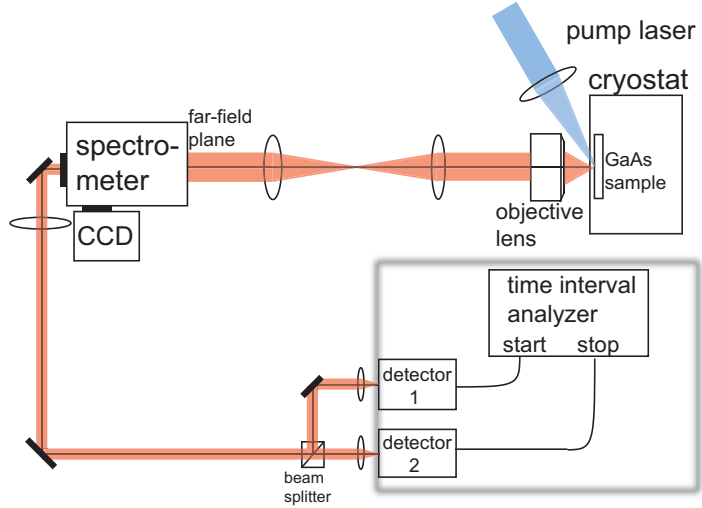}}
    \caption{(Color Online) Experimental setup for the dispersion relation
measurements. The blue beam is that of the pump laser, while the red
beam represents the PL from the sample. For time-resolved
spectroscopy, the spectrometer was replaced by a streak camera
attached to a monochromator.}\label{setup}
   \end{center}
  \end{figure}
The sample had a microcavity structure consisting of AlAs/AlGaAs
DBRs. To obtain the PL from the top surface, DBRs with 16 (20) top
(bottom) layers were used. The $12$ QWs were divided into three groups
and positioned at the three highest-mode-intensity antinodes of the
microcavity. This sample showed strong coupling at low temperatures,
and the normal-mode splitting at $k = 0$ was $14$ meV around zero
detuning. The pump laser was a mode-locked Ti:sapphire laser with a
$3$ ps pulse width; light was injected into the sample at an angle
of 50--$60^{\circ}$ from the normal, which corresponds to $k_{||}
\sim 7 \times 10^4$ cm$^{-1}$. The wavelength of the pump laser was
set to $747$ nm at $8$ K to maximize the rate of injection into the
sample reflection dip due to the cavity photon mode. As the
temperature was increased, the cavity photon energy experienced a
redshift, which meant that the pump-laser wavelength needed to be
changed with temperature.

A schematic of the experimental setup is shown in Fig.~\ref{setup}.
A total of three lenses including an objective lens with a large
numerical aperture of $0.55$ for large $\Delta k$ collection were
used for imaging the far-field plane on the entrance slit of the
spectrometer. The imaging of the far-field plane means that the radial
coordinate corresponds to the PL emission angle from the sample,
with the in-plane momentum of the polariton or the photon in the
sample being defined as $k_{||}=2\pi/\lambda \sin{\theta}$, where
$\lambda$ is the PL wavelength and $\theta$ is the PL emission
angle. The emission in the horizontal direction is deflected in the
spectrometer by a grating to create the energy axis. Since the
vertical axis corresponds to the in-plane momentum, the dispersion
relation is recorded by a CCD camera attached to one exit of the
spectrometer. The other exit of the spectrometer is used to send the
PL to a Hanbury Brown and Twiss (HBT) setup to measure the
second-order correlation function. The HBT setup consists of a beam
splitter, single-photon detectors (Perkin Elmer, SPCM-AQRH-14), and
a time-interval analyzer (ORTEC, picosecond time analyzer 9308). The
spectrometer grating is adjusted after every measurement at each
temperature to obtain the whole spectrum since the PL energy shifts
as the excitation density changes when the pump laser power changes.
The width of the spectrometer exit slit, corresponding to the width
of the observed energy, is set so as to obtain a sufficiently wide
spectrum with the HBT setup ($\sim \, 4$ nm). The detection width of the
in-plane momentum is $\Delta k \sim \, 2 \times 10^4$ cm$^{-1}$. When
we perform time-resolved spectroscopy, the spectrometer is replaced
by a streak camera with a $2$ ps time resolution. However, the time
resolution is degraded by the temporal dispersion of the grating
inside the monochromator attached to the front of the streak camera.

\subsection{Temperature dependence below threshold}
\begin{figure}
\begin{center}
\scalebox{0.17}{
\includegraphics{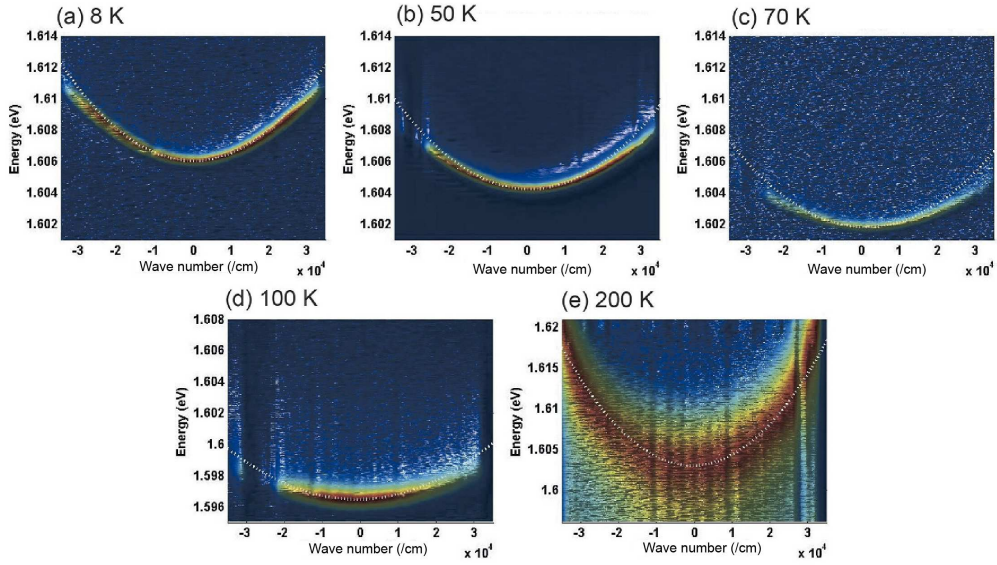}}
\caption{(Color Online) Dispersion relations below threshold at (a) $8$, (b) $50$, (c) $70$, and (d) $100$ K. 
The dispersion for the LP branch is shown. Cavity-mode branch dispersion was
observed at (e) $200$ K. As the temperature increased, the effective
mass became heavier and the detuning increased because the energy of
the exciton was redshifted as the temperature of the exciton increased faster than
the cavity temperature. At $200$ K, the effective
mass was the same as the mass of the cavity photon since strong
coupling no longer existed owing to the large detuning, and the
system was in the weak-coupling regime.}\label{dispersions}
   \end{center}
  \end{figure}
\begin{table}[ht]
\caption{Effective masses and detuning values at various temperatures. The
effective mass of the cavity was evaluated to be $ m_{\rm cav}=2.8
\times 10^{-35}$ kg.} \label{temp} \footnotesize\rm
\begin{tabular}{cccccc}
\hline
 Temperature (K) &8 & 50 & 70 & 100 & 200 \\
Effective mass ($m_{\rm cav}^{-1}$)&$2.5$  &$2.8$  &$2.9$  &$4.4$ &
$1$
\\
Detuning (meV)&$2.5$ &$3.8$ &$4.5$ &$9.0$  &-- \\ \hline
\end{tabular}
\end{table}
We began with an investigation of the temperature dependence of the
normal-mode dispersion relations seen in the low-excitation regime
below the condensation or lasing threshold. As the temperature
increased, we found that both the exciton energy and the cavity
photon energy were redshifted. The redshift of the exciton energy
occurred faster than that of the cavity photon energy.
Furthermore, the detuning $\Delta E(k_\parallel) =
E_{cav}(k_\parallel)-E_{exc}(k_\parallel)$, where $E_{\rm
cav}(k_\parallel)$ is the cavity photon energy and $E_{\rm
exc}(k_\parallel)$ is the exciton energy, increased as the system
temperature increased. We experimentally measured dispersion curves,
fitted a quadratic curve to the dispersion curve near
$k_{\parallel}=0$, and estimated the effective mass from the
curvature. In the case of the LP branch, a wider fitting region
resulted in a heavier evaluated effective mass (see Fig.~\ref{fit}
in Appendix~\ref{fitting} for the fitting region).

Dispersions taken below the threshold at various temperatures
obtained by time-integrated far-field spectroscopy are shown in
Fig.~\ref{dispersions}. These dispersions represent the
temperature-induced transition from the small-mass LP mode at 8 K to
heavier exciton-like heavier LP modes at higher temperatures of $50$, $70$,
and $100$ K, and then the further transition to the cavity mode at
$200$ K (the lasing threshold is reached in this mode). Note that
the lower branch changed from that of the LP to that of the exciton,
and so no threshold behavior was seen in the lower energy branch at
$200$ K.

The detuning values, listed in Table \ref{temp} along with the effective
masses, were evaluated from the cavity photon mass at $200$ K and
the effective mass of the LP. The refractive index of $3.14$ for the
cavity photon mass at $200$ K was lower than that of GaAs because of
the layered DBR structure containing Ga$_{0.8}$Al$_{0.2}$As and
AlAs, whose refractive indexes are smaller than that of GaAs. At $8$
K, the LPs relaxed to the bottom of the branch and formed a
polariton condensate when the pump power was increased. As the
temperature increased, the effective mass increased, as did the
detuning. The LP mode became exciton-like owing to the fast redshift
of the exciton energy, before separating into heavy hole (hh)
exciton and cavity photon modes, which is accompanied by a
transition from strong coupling to weak coupling. At $200$ K, the smaller
effective mass indicates the cavity photon mode.

This study differs from that in ref.~\citen{tempel2} is that
the sample position was not changed as the detuning was scanned in
response to the temperature change. This allowed us to screen out
unexpected effects due to position-dependent imperfections in the
sample and was done partly because the temperature dependence of
constant detuning has already been studied in ref.~\citen{tempel2}.
Since our aim was to investigate standard vertical cavity
surface-emitting laser (VCSEL)-type lasing, where there is no strong
coupling between an exciton and a cavity photon below the lasing
threshold at $200$ K, this high-temperature condition at which
exciton binding is difficult was suitable.

\subsection{Excitation power dependence}
In examining the dependence of the normal modes on the excitation
density at several temperatures, we found that the PL of the
high-density polariton condensates at a low temperature ($8$ K)
displayed distinct features in terms of the PL intensity, energy,
linewidth, and second-order correlation function compared with the
PL of the photon laser obtained at a high temperature ($200$ K).
Moreover, the temperature data ranging from $8$ to $200$ K clearly shows
how each feature of the polariton condensate converges to
each of those of the photon laser.
\subsubsection{PL intensity}
\begin{figure}
   \begin{center}
    \scalebox{0.17}{
    \includegraphics{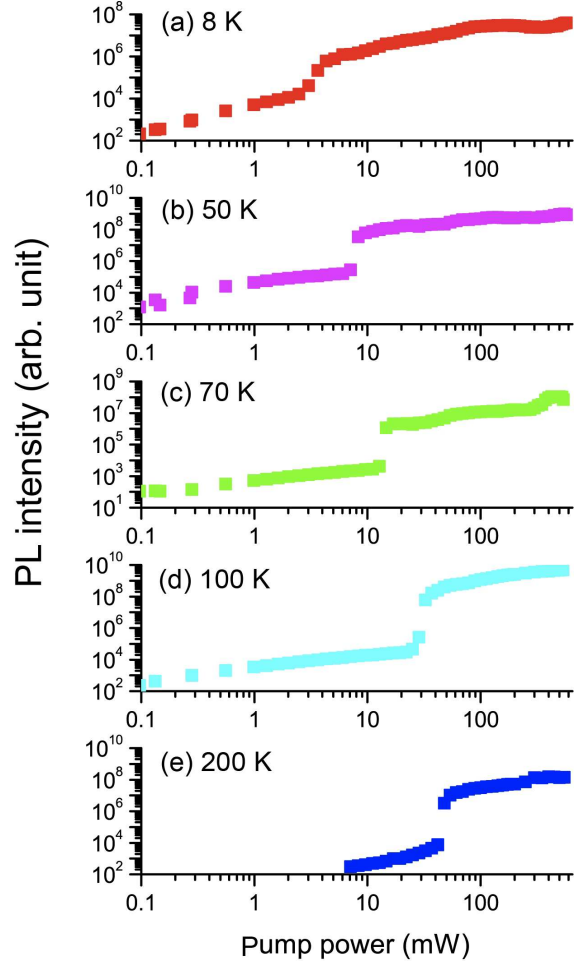}}
    \caption{(Color Online) Excitation power dependence of the PL intensity at $k=0$. (a)
$8$, (b) $50$, (c) $70$, (d) $100$, and (e) $200$ K. At $200$ K, 
the PL intensity was weak and undetectable below
$6$ mW.  }\label{intensity}
   \end{center}
  \end{figure}
The excitation power dependence of the PL intensity, shown as a
function of the pump laser power, for various temperatures is shown
in Fig.~\ref{intensity}. A nonlinear increase in the PL intensity is
seen for each temperature, and as the temperature was increased, the
threshold laser power, i.e., the threshold excitation density,
increased. A higher excitation density is needed to reach the lasing
threshold and therefore achieve population inversion, whereas this is not
necessary for the polariton system \cite{imamoglu}. Note that the
normal modes below the threshold for temperatures lower than $100$ K
were the LP and UP, and we can see that the LP branch reached the
threshold. By increasing the temperature, the UP gradually changes
to the cavity mode, and LPs become excitons with the flat curvature
of a heavy effective mass. Thus, the normal mode at $200$ K was the
cavity mode due to exciton dissociation and the large detuning, that is,
the cavity mode was already in the conduction band. The transition
of the cavity mode to the lasing mode occurred at $200$ K.

The lowest threshold pump power of approximately $3$ mW was recorded for
the $8$ K case [Fig.~\ref{intensity} (a)], and as the excitation
increased, a monotonic increase in the PL intensity was seen.
However, a distinct second threshold was not observed, unlike
in previous studies
\cite{dang,hui2,bajoni,nelsen,bajoni2,elena,tempel}. As expected
from BEC thermal equilibrium theory, the threshold increased with an
increase in the temperature but the second threshold was not
observed until $70$ K [Fig.~\ref{intensity} (c)], where a nonlinear
PL intensity increase was seen at around $300$ mW, implying a
transition to photon lasing. At $100$ K and $200$ K
[Figs.~\ref{intensity} (d) and (e)], the second threshold was not
observed, which implies that the system has already reached photon
lasing after the first threshold. However, as we will show later, the
properties of the second-order correlation function make this
slightly ambiguous in the case of $100$ K.

\subsubsection{PL energy}
\begin{figure}[Hh]
   \begin{center}
    \scalebox{0.17}{
    \includegraphics{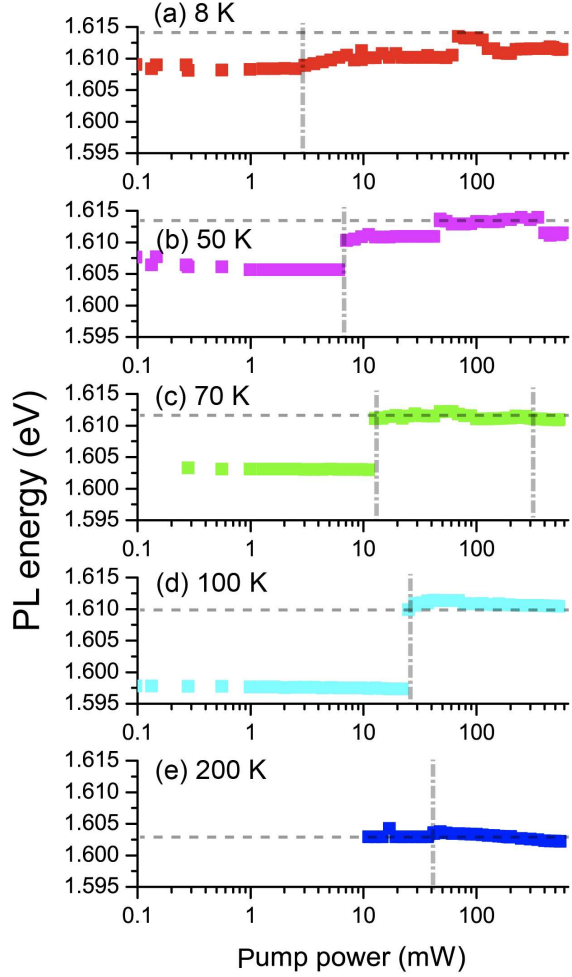}}
    \caption{(Color Online) Excitation power dependence of the PL energy. The dashed
    lines are the energy of a cavity photon of $k=0$ at each temperature. The
discrepancy between the dashed line and the PL energy at $100$ K is
due to the low precision in evaluating the cavity energy from the
effective mass. The limitation in the precision comes from the broad
linewidth due to the thermal energy. The dot-dashed lines correspond
to the thresholds in the PL intensity.
  }\label{energy}
   \end{center}
  \end{figure}
\begin{figure}[Hh]
   \begin{center}
    \scalebox{0.13}{
    \includegraphics{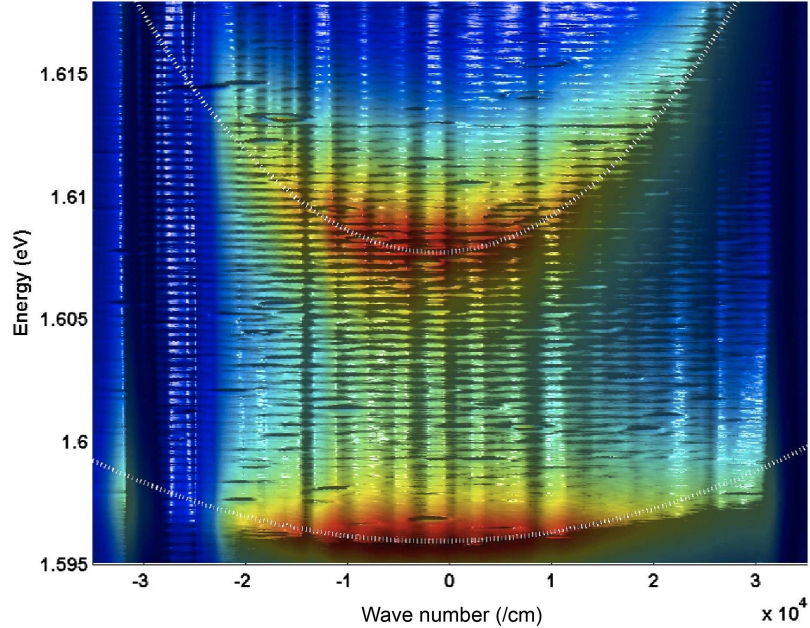}}
    \caption{(Color Online) Dispersion around the threshold at $100$ K.
  }\label{100K}
   \end{center}
  \end{figure}
Looking at the change in the PL energy at $k=0$, which gives the
maximum PL intensity at each temperature (Fig.~\ref{energy}), we see
that in the $8$ K case [Fig.~\ref{energy} (a)] there is a blue shift
in the energy below the condensation threshold ($\sim 1.608$ eV)
at approximately the condensation threshold ($\sim 3$ mW). The energy gradually approached
the cavity photon energy of $1.615$ eV as the pump power increased.
The behavior was also seen at $50$ K [Fig.~\ref{energy}
(b)], where the condensation threshold was around $6$ mW. Note that
the increase in temperature resulted in a redshift of the energy
below the threshold, but a large blue shift of $5$ meV occurred at
the threshold. This blue shift was much larger than that in the $8$ K
case because the large excitonic component at $50$ K resulted in a
large interaction energy. With further increase in the excitation
density, there is another jump in the PL energy at around $40$ mW to
an energy $7$ meV higher than that below the condensation
threshold and equal to the cavity photon energy. However, even under
these conditions, the lower-energy peak ($5$ meV higher than the
energy below the threshold) reappeared in the higher-excitation regime
at around $400$ mW; a similar behavior was observed in the data at $8$ K.
Owing to the pumping of the pulse laser, the polariton density
showed a pulsed profile on a time scale of $1$--$10$ ps.
The PL emitted at the maximum excitation density originates from the 
cavity phonon energy, and as the density decreased, the energy of 
the PL also decreased.

A different behavior was observed at $70$ K [Fig.~\ref{energy} (c)]. A
large blue shift of $7$ meV was seen at the first threshold at around
$10$ mW, but there was no large change in the energy in the high-excitation 
regime, which may be due to heating by the strong pump
laser. The system remained in the weak-coupling regime even when the
excitation density decreased with a pulsed excitation profile. At
the second threshold ($\sim 300$ mW), the energy corresponding to
the maximum PL intensity remained the same as the cavity photon
energy. However, the presence of this second threshold raises the
possibility that the behavior in the intermediate regime differs
from photon lasing.

The $100$ K case [Fig.~\ref{energy} (d)] showed a similar behavior to
the $70$ K case, except for the amount of blue shift at the first
threshold. At low temperatures, the normal-mode splitting was $14$
meV, and the cavity photon energy was midway between the UP and the LP.
However, the detuning at $100$ K was large, and the cavity photon
energy was very close to the UP energy. Therefore, the blue shift was
larger than those at lower temperatures and almost the same as that
of normal mode splitting. In Fig.~\ref{100K}, the curvature of the
lower energy mode reveals a heavy LP mass due to the large detuning
at $100$ K, while the upper energy mode shows the curvature of the
cavity photon mode, which is brighter around the threshold pump
power. The energy above the threshold is equal to the cavity
photon energy, implying photon lasing.

\subsubsection{PL linewidth}
\begin{figure}[h]
   \begin{center}
    \scalebox{0.17}{
    \includegraphics{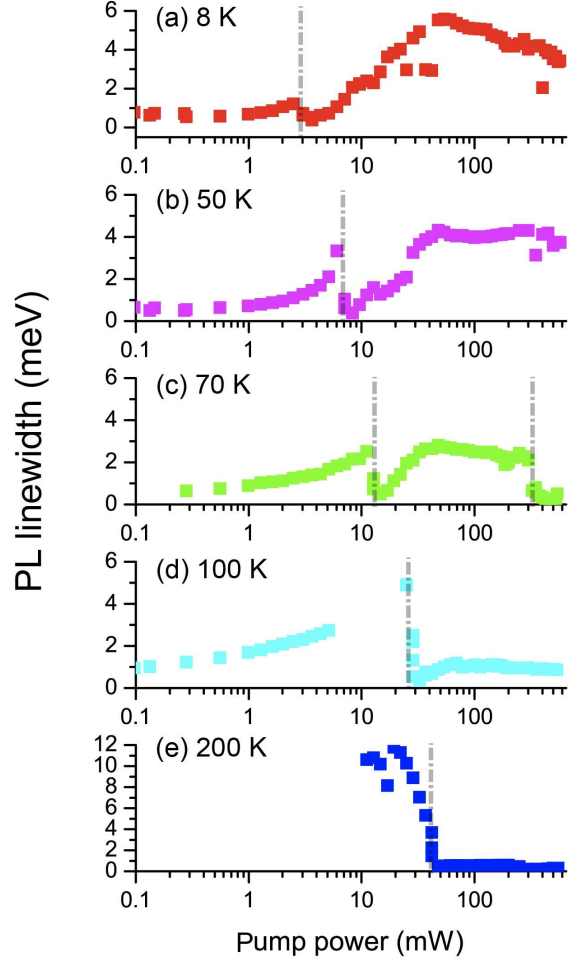}}
    \caption{(Color Online) Excitation power dependence of the PL linewidth. The dot-dashed
lines correspond to the thresholds in the PL intensity.
  }\label{linewidth}
   \end{center}
  \end{figure}
Turning to the linewidth, narrowing at the threshold was observed in
the $8$ K case [Fig.~\ref{linewidth} (a)], but this was followed by
broadening as the density increased owing to both
polariton--polariton interactions \cite{porras} and pulsed
excitation since the amount of blue shift depends on the excitation
density and the PL from various densities was integrated at the CCD.
The gradual narrowing of the linewidth in the high-density regime
shows that there are a possible concentration of the cavity photon
energy and a decrease in polariton--polariton interaction, which
indicate an approach to photon lasing. This trend was repeated at
$50$ K [Fig.~\ref{linewidth} (b)], but when the maximum PL energy
was equal to the cavity photon energy at around $40$ mW, no further
broadening occurred, with the linewidth being around $4$ meV, as in
the $8$ K case. The broad linewidth is a result of the PL from not
only the cavity photon energy but also the lower energy, as
discussed in the previous subsection.

Although linewidth narrowing at the threshold and the subsequent broadening
was also observed at $70$ K [Fig.~\ref{linewidth} (c)], the maximum
linewidth was less than $3$ meV. The narrower linewidth is due to
the fact that there was only one peak at the cavity photon energy.
However, this linewidth was still larger than that in the photon
lasing case [Fig.~\ref{linewidth} (e)] and implies 
that a polariton--polariton interaction effect remains. At the
second threshold ($300$ mW), narrowing due to the PL concentrating
to the cavity photon energy was observed, and a clear transition to
photon lasing occurred. At $100$ K [Fig.~\ref{linewidth} (d)],
narrowing occurred at the threshold; however, the subsequent
broadening was smaller than that at $70$ K because of a decrease in
the polariton--polariton interaction effect. At $200$ K
[Fig.~\ref{linewidth} (e)], the broad linewidth below the threshold
caused by the high thermal energy narrowed drastically at the lasing
threshold. The PL energy in this case was equal to the
cavity photon energy since there was no polariton--polariton
interaction effect, and hence, no linewidth broadening occurred
above the lasing threshold.

Although a narrow linewidth was observed at temperatures above $70$ K, in
the low-temperature cases ($8$ and $50$ K), the broad linewidth in
the higher excitation regime implies that polariton--polariton
interactions maintain their effect even in a high-density
regime on the order of a hundred times the condensation threshold.

\subsubsection{Second-order correlation function}
\begin{figure}
   \begin{center}
    \scalebox{0.17}{
    \includegraphics{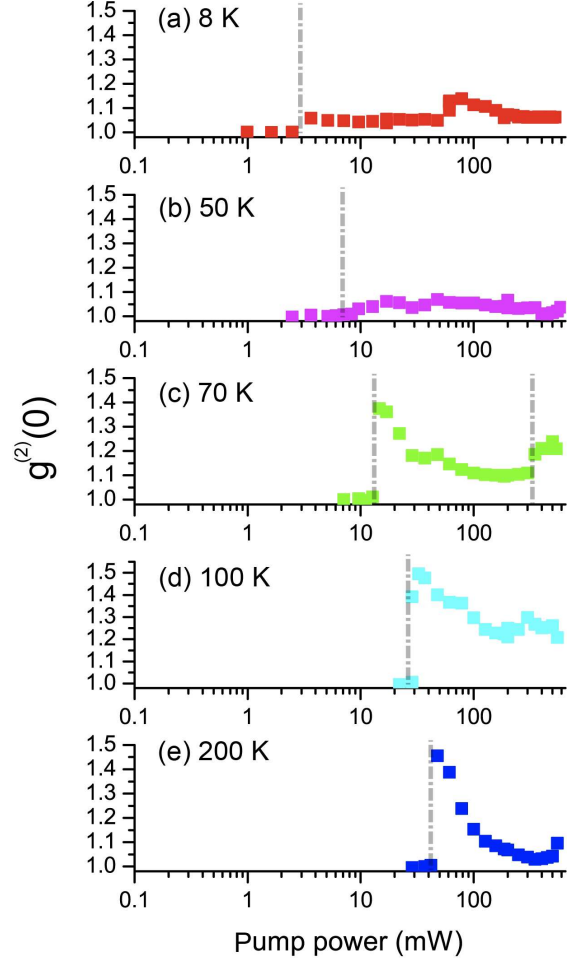}}
    \caption{(Color Online) Excitation power dependence of the second-order coherence
function. The dot-dashed lines correspond to the thresholds in the
PL intensity.
  }\label{g2}
   \end{center}
\end{figure}
We also examined the second-order correlation function
(Fig.~\ref{g2}) using the HBT setup; however, difficulties in
detecting weak signals prevented us from collecting data at low pump
powers. In the $8$ K case [Fig.~\ref{g2} (a)], the PL showed bunching
behavior above the threshold ($g^{(2)}(0)>1$)
\cite{hui,lausanne,horikiri,tempel,tempel2} up until the high-excitation-density regime, 
which was more than a hundred times the
condensation threshold ($>$300 mW). When the measurements were
performed using a high-time-resolution streak camera \cite{tempel},
the $g^{(2)}(0)$ below the threshold was closer to 2 after
the statistics of the thermal state were correctly obtained. Note
that, in this study, the time resolution of the photon detectors in
the HBT setup ($\sim 300$ ps) was not sufficient to resolve the
signal in one pulse. Average statistics were obtained because the
intensity correlation time may be much shorter than the PL lifetime:
we did not observe any correlation between the photons at the
beginning and end of the PL pulse \cite{hui}. The results for
$g^{(2)}(0)$ were thus equal to unity.

When the system is in the condensation regime, the PL pulse width is
reduced by more than one order of magnitude
\cite{horikiri,tempel,tempel2}. The intensity correlation time of
the system covers or is closer to the entire PL pulse.
Therefore, even though the time resolution of the detectors is
greater than the PL lifetime, the detection system correctly collects
the photon statistics \cite{lausanne,horikiri}. If the actual
correlation time is shorter than the pulse duration, then the
$g^{(2)}(0)$ obtained underestimates its true value
\cite{lausanne}, and in fact, even if the measured results do not
reflect the true statistics, they still indicate the bunching and
the difference from photon lasing that occurs far above the threshold
(see Appendix~\ref{time-integration} for more details).

Bunching in the broad excitation density regime was also seen at
$50$ K [Fig.~\ref{g2} (b)], but the strong bunching above the
threshold at $70$ K [Fig.~\ref{g2} (c)] was different from that
at low temperatures. The behavior at these two temperatures was
similar to that in the case of photon lasing. $g^{(2)}(0)$ did
not reach unity but rather increased after the second threshold,
as also observed in a previous study \cite{tempel}. At $100$ K
[Fig.~\ref{g2} (d)], the absence of a clear second threshold in the
PL intensity meant that there was also no clear second threshold in
$g^{(2)}(0)$; however a clear convergence to unity was not seen
at this temperature, implying a difference from photon lasing
although an even higher density might cause convergence. In the
$200$ K case [Fig.~\ref{g2} (e)], $g^{(2)}(0)$ monotonically
decreased to unity after the transition at the threshold, which is a
typical characteristic of photon lasing \cite{ulrich}.

\section{Discussion}\label{discussion}
In the previous section, we have reported several PL characteristics 
at various temperatures. In this section, we discuss the lasing mechanism 
at each the temperature. 

At $200$ K, there are no e--h correlations from 
the observation of the dispersion curve below threshold 
[Fig. \ref{dispersions} (e)]. The e--h plasma is formed, increasing the 
excitation power. Owing to the same microcavity structure as the VCSEL, 
conventional photon lasing occurs from the observations of 
the same energy of the PL peak to the cavity energy [Fig. \ref{energy} (e)], 
the narrow linewidth 
[Fig. \ref{linewidth} (e)], and the converged 
behavior of the second-order autocorrelation function [Fig. \ref{g2} (e)]
above the first threshold. While the phenomena observed at $200$ K 
is similar to those observed at $100$ K, particularly, the narrow PL linewidth 
after the first threshold in Fig. \ref{linewidth} (d)), the two PL 
peaks are observed in Fig.~\ref{100K}.
This fact cannot be explained in the standard photon lasing. Therefore, 
we cannot claim that this threshold is identified to the transition 
to the standard photon lasing. 

At $8$ K, we observe the blue shift behavior after the first threshold 
[Fig. \ref{energy} (a)]. This can be explained as follows: This condensation is taken as 
the exciton-polariton BEC, which is the weakly-interacting BEC. However, 
increasing the excitation power, we still observe the PL characteristics similar to those above the first threshold while the system 
reaches to the nonequilibrium situation. We report that the second threshold is not 
observed. In the following subsection, we discuss the 
evaluation of the excitation density of the e--h pair. At $50$ K, we also 
observe the PL characteristics similar to those observed at $8$ K 
after the first threshold in the blue shift [Fig. \ref{energy} (b)], 
the increasing linewidth [Fig. \ref{linewidth} (b)], and the almost 
constant second-order autocorrelation function (Fig. \ref{g2} (b)). 
Note that, the effect of all PL characteristics at $50$ K is weaker 
than that in the $8$ K case. However, at $70$ K, we observe that 
the PL energy rapidly reaches to the cavity energy; however, we did not observe 
the blue shift behavior after the first threshold [Fig. \ref{energy} (c)], 
which is completely different from those in the $8$ and $50$ K cases. Since the PL linewidth 
does not be narrow after the first threshold as in $100$ and $200$ K cases, 
this transition cannot correspond to the standard photon lasing. Therefore, 
this observation may be due to the dynamical condensation of the 
exciton-polariton system. Thereafter, the same PL characteristics as those in  
the standard photon lasing is measured after the second threshold. 

\begin{table}[t]
\begin{center}
\caption{Phase transitions at various temperatures.}
\footnotesize\rm
\begin{tabular}{ccc} \hline
Temperature (K) & First threshold & Second threshold \\ \hline 8 &
polariton condensation & \\ 50 & polariton condensation &
\\ 70 & polariton condensation & photon lasing \\ 100 & photon lasing? & \\
200 & photon lasing & \\ \hline
\end{tabular}
\label{final_results}
\end{center}
\end{table}
The above discussions are summarized in Table~\ref{final_results} to relate 
our experimental data with our previous knowledge of the exciton-polariton system. 
While we have not directly measured the e--h pair or the e--h correlations 
as alluded previously, we can evaluate the excitation density of the e--h pair. Note that 
the estimated excitation density of the e--h pair is not always the same 
as the actual density of the e--h pair since the e--h pair may break. 

\subsection{Estimation of excitation density of the exciton-hole pair}
To determine whether our system reaches the semiconductor inversion
condition, i.e., the Bernard--Duraffourg condition~\cite{elena}, we
evaluated the excitation density using a picosecond mode-locked
laser with a $76$ MHz repetition rate. By estimating the bandwidth
as well as the reflection dip at the energy of the pump laser, we
could estimate the input rate as $\sim \, 0.1$. Considering the
absorption coefficient of the QWs at the pumping energy ($\sim \, 2
\times 10^4$ cm$^{-1}$), the absorption rate at each QW is around
$0.014$. The absorption probability from an input of one pump photon
is $1.4\times 10^{-3}$, and there are around $2.5 \times 10^{10}$
photons per pulse at an average power of $500$ mW. Thus, the
excitation density of a spot roughly $50\times 100$ $\mu$m in size
is $1.4\times 10^{-3}\times 2.5\times 10^{10}/(50\times 100\;{\rm
\mu m}^2)=7 \times 10^{11}$ cm$^{-2}$, which is close to that
derived for the pulsed condition given in ref.~\citen{elena}. In
ref.~\citen{elena}, the condition was estimated to be close to the
Bernard--Duraffourg condition, but the estimated excitation density
had a large uncertainty owing to the experimental conditions used. For
example, the input rate of $0.1$ was estimated roughly and not
determined precisely from measurements, as was the absorption rate.
Moreover, temperature determination is always a problem in this
field. However, the possibility of the system being close to the
inversion condition is implicitly implied in the high-excitation
regime, and if the inversion condition is satisfied, then this leads
to the new type of lasing discussed in the high excitation regime
and theoretically studied in the nonequilibrium condition
\cite{yamaguchi,yamaguchi2} using the framework developed in
ref.~\citen{szymanska}. Although we have not obtained direct
evidence of the e--h pair correlation in the high-excitation regime,
recent observations of a ghost polariton branch \cite{kohnle,zajac}
using an intense resonant excitation laser indicate the possibility
of e--h pairs and polaritons even in the high-density regime.

\section{Conclusion \label{conclusion}}
We observe the intensity, the energy of the main peak, the linewidth, and 
the second-order autocorrelation function of the PL from an exciton--polariton
system at various temperatures and excitation powers. 
Our observation is summarized in Table \ref{final_results} from the viewpoint of 
the nonlinear gains of the PL intensity. 
Contrary to conventional expectations, we did not observe the 
second threshold at a low-temperature, high-excitation-density regime. 
The characteristics of the low temperature 
data differ from those of conventional photon lasing observed at high 
temperatures. Since we have evaluated the excitation density 
of e--h pairs in an experimental setup that satisfies the 
Bernard--Duraffourg condition
at low temperatures, that is, the high-excitation-density regime, 
the lasing mechanism of the exciton--polariton system
is different from that of a standard photon lasing system.
Our results suggest that e--h binding is present in the 
high-excitation regime at low temperatures, supporting the 
theoretical considerations~\cite{kamide1,tim,kamide2,yamaguchi,yamaguchi2}.

\section*{Acknowledgments}
The authors wish to thank T. Ogawa, T. Byrnes, K. Kamide, M.
Yamaguchi, and N. Ishida for their helpful comments. This research
was supported by the Japan Society for the Promotion of Science
(JSPS) through its FIRST Program and KAKENHI Grant Numbers 24740277
and 25800181, a Space and Naval Warfare Systems (SPAWAR) Grant
N66001-09-1-2024, the Ministry of Education, Culture, Sports,
Science and Technology (MEXT), the State of Bavaria, the National
Institute of Information and Communications Technology (NICT), and
the joint studies program at the Institute for Molecular Science.

\appendix
\section{Parameters for the Estimation of Effective Masses} \label{fitting}
\begin{figure}
  \begin{center}
    \scalebox{0.15}{\includegraphics{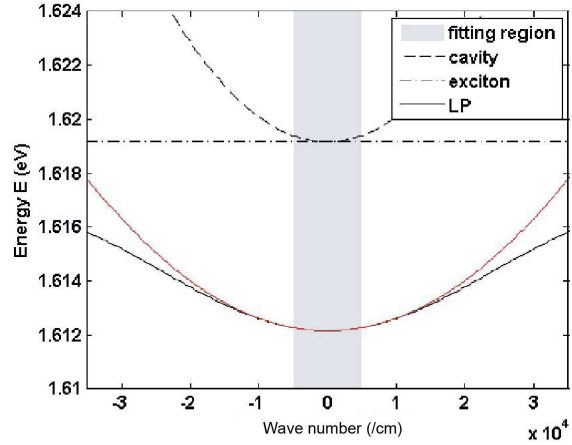}}
    \caption{(Color Online) Evaluation of the accuracy of the fitting of
$E_{LP}(k_\parallel) =
1/2\bigr[E_{exc}(k_\parallel)+E_{cav}(k_\parallel)-
\sqrt{4\hbar^2\Omega^2+(E_{exc}(k_\parallel)-E_{cav}(k_\parallel))^2}\bigl]$
(solid black curve) by a quadratic curve (red curve). The fit used $29$
data points in the region $|k_{||}|<0.5 \times 10^4$ cm$^{-1}$. The
effective mass obtained from the fit was $7.48 \times 10^{-35}$ kg
with a relative error of $0.26 \%.$ From this evaluation,
experimental data points in the region $|k_\parallel|<0.5 \times
10^4$ cm$^{-1}$ were analyzed to evaluate the effective
mass.}\label{fit}
  \end{center}
 \end{figure}

The effective mass of the LP is described by
$1/m_{LP}=|X|^2/m_{exc}+|C|^2/m_{cav}$, where $|X|^2$ and $|C|^2$
are the Hopfield coefficients determined by the detuning and normal-mode 
splitting of $2\hbar\Omega$, respectively. Here, the effective mass of the
exciton, $m_{exc}$, is defined by $1/m_{exc}=1/m_{e}+1/m_{hh}$ in
terms of the electron mass $m_e$, the heavy hole (hh) mass $m_{hh}$, and
the effective mass of the cavity photon $m_{cav}$. From the dispersion
relations, the effective mass $m^*$ is calculated as
$1/m^*=(1/\hbar^2)d^2E(k)/dk^2$. In the range of $k_\parallel\ll
k_\perp$, the energy of the cavity mass $E_{cav}=\hbar c/n_c
({k_\parallel}^2+{k_\perp}^2)^{1/2}$ is approximated by a quadratic
curve, and thus, the effective mass at $k_\parallel=0$ is $m_{cav} =
2\pi\hbar{n_c}^2/(\lambda_c c)$ for the refractive index $n_c$ and the
resonant frequency $\lambda_c$. The effective mass at the bottom of
the LP branch is 7.46 $\times10^{-35}$ kg. The following parameters
were used to estimate the effective masses: $2\hbar\Omega=14$ meV, a
free electron mass of $m_0 = 9.109 \times 10^{-31}$ kg, $m_e=0.0665
m_0$, $m_{hh}=0.111 m_e$, $\lambda_c=767.5$ nm, $n_c=3.6$ for GaAs,
and $\Delta E =0$. The fitting region is shown in Fig.~\ref{fit}.

\section{Effects of Time Integration on the Linewidth Measurements} \label{time-integration}
\begin{figure}[th]
   \begin{center}
    \scalebox{0.32}{
    \includegraphics{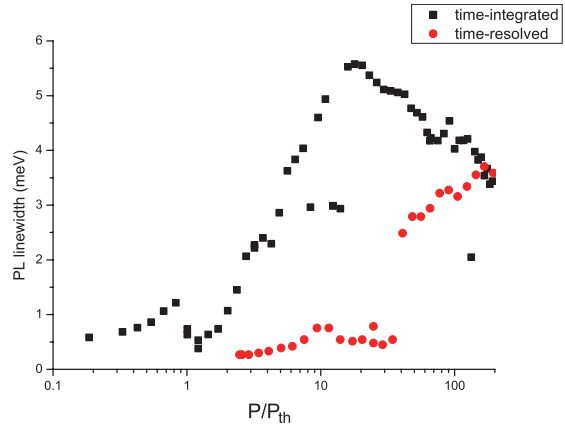}}
    \caption{(Color Online) Excitation power dependence of the PL linewidth at $8$ K.
Time-integrated spectroscopy data are shown by black squares, while
time-resolved spectroscopy data are shown by red circles. The
time-resolved data are extracted at the time corresponding to that of the
maximum PL intensity. }\label{linewidth2}
   \end{center}
  \end{figure}
At a low temperature ($8$ K), time-resolved linewidth data
were also collected using a streak camera (Fig.~\ref{linewidth2}).
In contrast to the time-integrated data, the time-resolved data
showed a narrower linewidth, especially in the relatively low
excitation regime. These data reflect the fact that the degree of
blue shift depends on the excitation density and the profile of the
pulse pump laser. When the time-varying excitation density was
summed, the observed linewidth became broader owing to the sum over
the various densities. However, as the excitation increased, the
time-resolved data also showed a broad linewidth at the time of
maximum PL intensity. In the maximum excitation regime, the
linewidth was about the same as that for the time-integrated data,
which demonstrates that the linewidth was much broader than those in the
cases of higher temperatures ($70$, $100$, and $200$ K) corresponding
to the highest excitation regimes. The time resolution of streak
camera spectroscopy was better than $10$ ps owing to the temporal
dispersion imposed at the grating of the monochromator attached to
the streak camera \cite{elena}. Note also that spatial
inhomogeneities can also increase the linewidth, although this
effect would appear to be small since we collected the emission near
$k=0$ where the intense emission is dominated by the central area of
the pump spot. Nevertheless, the difference between the
time-integrated and time-resolved measurements in
Fig.~\ref{linewidth2} reveals that the polariton linewidth of the
time-integrated data broadens the obtained results. We also see that
this broad linewidth in the relatively low excitation regime is due
to time integration rather than to polariton--polariton
interactions. However, in the high-excitation regime, it is
possible that the broad linewidth is a result of
polariton--polariton interactions.

\end{document}